\documentclass[conference]{IEEEtran}
\usepackage{cite}
\usepackage{amsmath,amssymb,amsfonts}
\usepackage{algorithmic}
\usepackage{graphicx}
\usepackage{textcomp}
\usepackage{xcolor}
\def\BibTeX{{\rm B\kern-.05em{\sc i\kern-.025em b}\kern-.08em
    T\kern-.1667em\lower.7ex\hbox{E}\kern-.125emX}}

\newcommand{\STAB}[1]{\begin{tabular}{@{}c@{}}#1\end{tabular}}

\newcommand{\eq}[1]{ \begin{align}#1\end{align}}

\DeclareMathOperator*{\argmax}{arg\,max}  

\graphicspath{{./figs/}}

\usepackage{booktabs}
\usepackage{rotating}
\usepackage{makecell, multirow}

\begin{document}

\title{Detecting, identifying, and localizing radiological material in urban environments using scan statistics} 


\author{\IEEEauthorblockN{Michael D. Porter, Alphonse Akakpo \\ 
}
\IEEEauthorblockA{University of Virginia\\ Charlottesville, VA\\
Email: \{mdp2u, ana2cy\}@virginia.edu}
}

\maketitle

\begin{abstract}
A method is proposed, based on scan statistics, to detect, identify, and localize illicit radiological material using mobile sensors in an urban environment. Our method handles varying levels of background radiation that change according to an (unknown) environment. Our method can accurately determine if a source is present along a street segment as well as identify which of six possible sources generated the radiation. Our method can also localize the source, when detected, to within a few seconds. We have presented our results across a range of decision thresholds allowing stakeholders to evaluate the performance at different false alarm rates. 
Due to the simplicity of our approach, our models can be trained in a few minutes with very little training data and holds the potential to score a run in real-time.  
Our method was one of the top performing submissions in the \emph{Detecting Radiological Threats in Urban Areas} competition.
\end{abstract}


\section{Introduction}
The machine learning challenge \emph{Detecting Radiological Threats in Urban Areas} is concerned with detecting, identifying, and locating radiological material using a moving sensor in an urban environment. See the website \cite{topcoder} for full details. The scenario is a mobile sensor that moves in a straight line on a simulated road measuring gamma-ray energy levels. If there is a radiological source present along the run, it emits energy according its energy spectra. The energy received at the sensor is obfuscated by the presence of background radiation sources and reflections from the urban environment. 
For a set of testing runs, the objectives are to: \textbf{detect} if there is a radiological source in the run, \textbf{identify} the source (six possible sources each with shielded or non-shielded profiles), and \textbf{locate} the time at which the sensor was closest to the source.

\section{Data and Challenge Details}

The National Nuclear Security Administration (NNSA) has teamed with Topcoder to host a radiological detection competition \cite{topcoder}. 
The competition organizers have used a Monte Carlo particle transport model to generate data from thousands of runs that mimic what would be acquired by a mobile $2 \times 4 \times 16$ NaI(Tl) detector moving down a simplified street in a mid-sized U.S. city. 

The runs are configured to capture the variability in an urban setting; the sensor can move down a fictitious street in different lanes and travel directions, move at different speeds (between 1.0 and 13.4 meters per second), and encounter different street geometries (e.g., building layouts, construction material, land-use, cross streets, etc.). 
Runs with a radiological source will vary by the type of source (including shielded and unshielded versions), the strength of source, and the location of the source along the street. 
The source types are provided in Table~\ref{tab:SourceID}.

\begin{table}[h]
    \centering
    \caption{Description of radiological sources}
    \label{tab:SourceID}    
    \begin{tabular}{ccl}
SourceID & Source Name & Source Description  \\
\toprule
0  &    Null &     No Source \\
1  &    HEU & Highly enriched uranium \\
2  &    WGPu & Weapons grade plutonium \\
3  &    131I & Iodine, a medical isotope \\
4  &    60Co & Cobalt, an industrial isotope \\
5  &   99mTc & Technetium, a medical isotope \\
6  &  99mTc+HEU & A combination of 99mTc and HEU
    \end{tabular}
\end{table}

A total of 9700 labeled runs were made available (the data can be accessed at the contest website \cite{topcoder}). 
There are 4900 null runs and 800 runs from each of the six sources. 
The labels for each run indicated the radiological source and time the sensor was closest to the source (if non-null). 
The labels do not indicate if the sources were shielded. 
The run data only include the time and energy level recorded by the sensor. The sensor speed, sensor location, and other environmental factors were not provided. 

\begin{figure}
    \includegraphics[width=.45\textwidth]{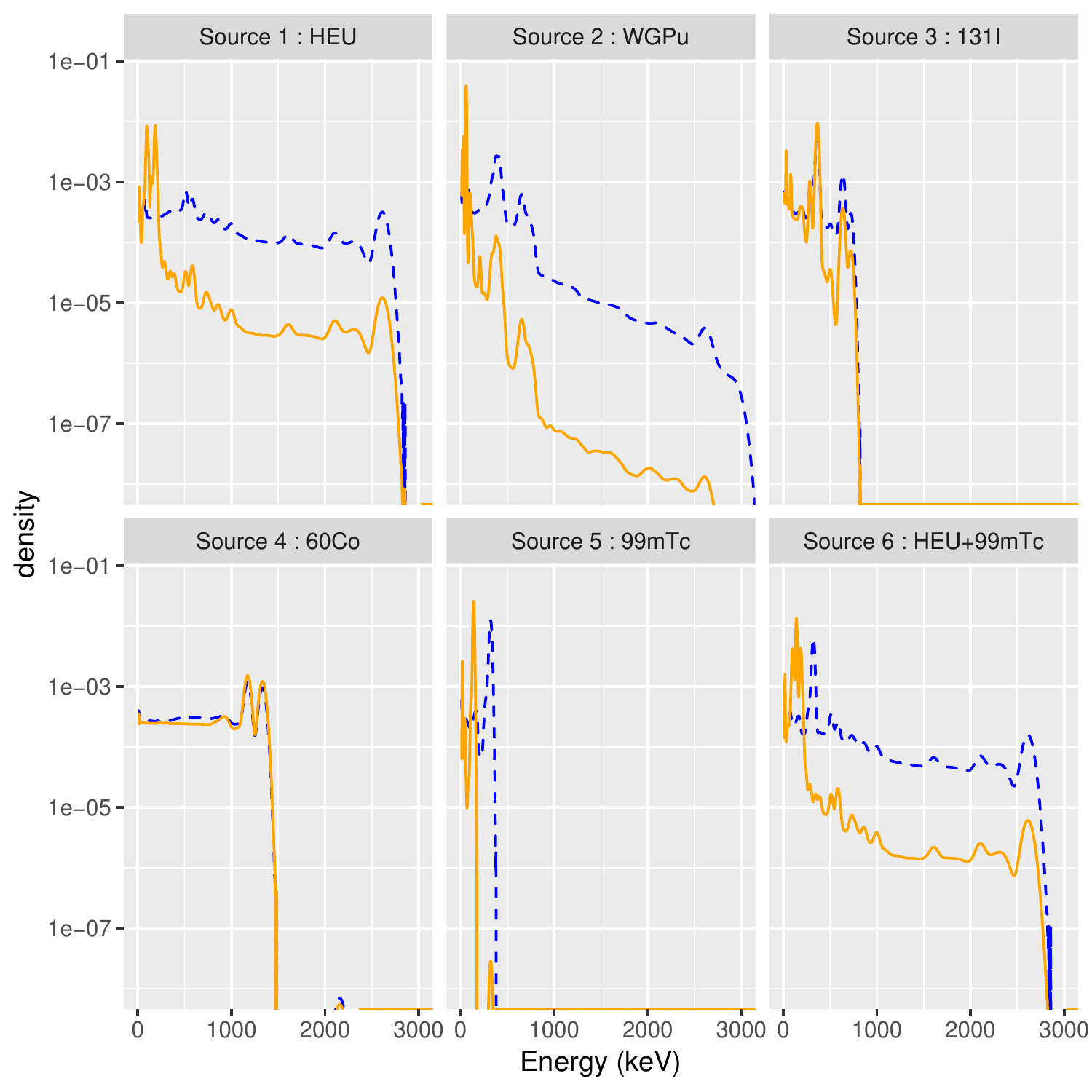}
    \caption{Estimated source densities. The solid orange line is for unshielded and the dotted blue line is for shielded sources. See Section \ref{Sec:estimation} for details of the estimation process.}
    \label{fig:pmf}
\end{figure}

In addition to the run data, the energy spectra for each source type (both shielded and unshielded) located 1 meter away from the sensor in a vacuum was also provided. From these, we estimated the density (standardized frequency) of the energy generated by each source (see Figure~\ref{fig:pmf}). 

\section{Methodology}

Our approach to detection, identification, and localization is based on the concept of scan statistics; that is, we search over all possible sources, locations, and emission profiles to obtain a test statistic that represents the combination that provides the strongest evidence for a radiological emitter. If the scan statistic exceeds a threshold, we declare that a radiological source is present. For any run in which a radiological source is suspected, our scan statistic approach will explicitly incorporate an estimate of the source type and source location. 

\subsection{Overview}
Let $k$ represent the source type, where $k=0$ implies no source and $k\in K=\{1.0, 1.1, 2.0, 2.1, \ldots, 6.0, 6.1\}$ corresponds to one of the 6 radiological sources with ($x.1$) and without ($x.0$) shielding. Let $\tau$ be the time (in secs) when the sensor is closest to the source. For each run, $\mathcal{H}_0$ is the null hypothesis of no source (i.e., $k=0$) and $\mathcal{H}(k, \tau)$ (for $k \in K$ and $\tau \geq 30$) is the hypothesis that the source is $k$ and the sensor is closest to the source at time $\tau$. The restriction $\tau \geq 30$ arises because the simulated contest data does not contain a source within the first 30 seconds of a run. 

A run is represented by the data $\{(t_i, x_i)\}_{i=1}^n$ where $t_i$ is the $i$th time and $x_i$ the $i$th energy received by the sensor since the start of the run. 
For a given run, the likelihood ratio (assuming independence between observations conditional on the model parameters) is 
\begin{align} 
\Lambda(k, \tau) &= \frac{\Pr(Data \mid \mathcal{H}(k,\tau))}{\Pr(Data \mid \mathcal{H}_0)} \\
&= \prod_{i=1}^n \frac{f(x_i \mid \mathcal{H}(k,\tau) )}{f(x_i \mid \mathcal{H}_0 )}  
\label{eq:LR}
\end{align}
which is the product of density ratios at the observations $x_1, \ldots, x_n$. 
The null density, $f(x_i \mid \mathcal{H}_0)$, can be directly estimated from the training data (see Section~\ref{Sec:estimation} for details). We assume that the density is stationary over all runs with no source. 
However the density $f(x_i \mid \mathcal{H}(k,\tau) )$, which is the density of observing an energy $x_i$, at time $t_i$, given that a source of type $k$ is located at time $\tau$, will depend on how close the sensor is from the source as well as the (unspecified in the contest data) factors that affect how much source radiation is received at the sensor (e.g., due to street geometry, building types, land use, etc.).

Given a source of type $k$ located at time $\tau$, we denote $\pi_{ik}(\tau)$ as the probability that an observation received at time $t_i$ came from the source (as opposed to the background radiation). This probability will be a function of the distance, $t_i - \tau$, the sensor is from the source and the source strength relative to the background radiation. 
A formal way to model the density $f(x_i \mid \mathcal{H}(k,\tau))$ is a statistical mixture 
\eq{
f(x_i \mid \mathcal{H}(k,\tau)) = f_k(x_i) \pi_{ik}(\tau) + f_0(x_i) (1-\pi_{ik}(\tau))
}
where $f_k(x_i)$ is the density of energy from source $k$, $f_0(x_i)$ is the density from the background, and $1-\pi_{ik}(\tau)$ is the probability that an observation at time $t_i$ would receive a measurement from the background. Plugging this into the density ratios of \eqref{eq:LR} gives
\eq{
\frac{f(x_i \mid \mathcal{H}(k,\tau))}{f(x_i \mid \mathcal{H}_0)} &= 
\frac{f_k(x_i) \pi_{ik}(\tau) + f_0(x_i) (1-\pi_{ik}(\tau))}{f_0(x_i)} \\
&= \frac{f_k(x_i)}{f_0(x_i)}  \pi_{ik}(\tau) + (1-\pi_{ik}(\tau)) 
\label{eq:mixture}
} 
This mixture model for $f(x \mid \mathcal{H}(k,\tau))$ could be estimated, for example, with the EM algorithm that incorporates prior information on $\pi_{ik}(\tau)$ according to knowledge about how far a source will emit radiation. However this information was not provided in the contest.

Because of the lack of information available to estimate $\pi_{ik}(\tau)$ we based our test statistics on a simpler representation of the strength of evidence that an observation arises from an extraneous source. 
Consider a single observation $x_i$ which is the recorded energy at time $t_i$. If this observation actually came from source $k$, then the log ratio
\eq{
r_{ik} = \log \frac{f_k(x_i)}{f_0(x_i)},
}
is expected to be greater than zero. However if it came from the background (i.e., $\mathcal{H}_0$), then $r_{ik}$ expected to be less than zero.
To reduce variability and focus attention on the observations that are more likely to be seen from the source, we use the statistic
\eq{
R_{ik} = \max\{0, r_{ik}\}
\label{eq:R}
}
as our measure of evidence that observation $i$ was from source $k$. 
Compared to the (log of) \eqref{eq:mixture}, \eqref{eq:R} only incorporates evidence in favor of an extraneous source, the lower threshold at 0 reduces the variability, and doesn't require estimation of $\pi_{ik}$. 

Building off this statistic, we estimate the log of the likelihood ratios in \eqref{eq:LR} as
\eq{
S(k, \tau, h) &= \log \left( \hat{\Lambda}(k, \tau) \right) \\
&= \sum_{i=1}^n K(t_i - \tau; h)  R_{ik} 
}
which is in the form of a kernel smoother where $K(t_i - \tau; h)$ is a Gaussian kernel with bandwidth (standard deviation) $h \in H$. 
We considered bandwidths of $H = \{0.5, 0.75, 1, 1.25, 1.5\}$ seconds.
The idea is that if the source is located at time $\tau$, then it will produce the most observations at times close to $\tau$ and gradually produce fewer observations as the distance between the sensor and source increases. Thus the $R$'s are weighted according to how far they are away from the hypothesized $\tau$. 

Assuming that the source would emit radiation symmetrically along the run (i.e., ignoring the influence of buildings and street geometries since that information was not available), 
\eq{
\hat{\tau}_k = \argmax_\tau S(k, \tau, h)
\label{eq:tau_k}
}
is the estimated time when the sensor is closest to the source and the score
\eq{
\tilde{S}(k, h) &= \max_\tau S(k, \tau, h) \\
&= S(k, \hat{\tau}_k, h)
}
is our measure of the evidence that radiation is being emitted from a source of type $k$ (using bandwidth $h$). 

The next steps are to determine if $\mathcal{H}_0$ should be rejected and, if so, estimate which alternative hypothesis is correct. One problem with using $\tilde{S}(k, h)$ as the test statistic is that its variance may differ over $k$ and $h$. Thus, if the variance is not accounted for the model may produce too many false alarms and over-attribute to certain sources.
To help rectify this, we estimated the mean, $\mu_0(k,h)$, and standard deviation, $\sigma_0(k,h)$, of $\tilde{S}(k, h)$ when the data are generated under $\mathcal{H}_0$. The standardized score, 
\eq{
Z(k,h) = \frac{\tilde{S}(k,h) - \mu_0(k,h)}{\sigma_0(k,h)} 
}
will have approximately the same mean and variance over all $k$ and $h$ allowing it to provide a fairer comparison between sources and over bandwidths. 

Our scan statistic for a run is the maximum value of $Z$ over all options for sources and bandwidths, 
\eq{
T = \max_{k\in K,\, h \in H} Z(k,h).
\label{eq:scanstat}
}
We reject the null of no source if $T \geq \phi$, for some threshold $\phi$, and estimate the source as
\eq{
\hat{k} = \argmax_{k \in K} \max_h Z(k, h)
}
and source location as
\eq{
\hat{\tau} = \hat{\tau}_{\hat{k}}
}
where $\hat{\tau}_{k}$ is given in \eqref{eq:tau_k}.

\subsection{Parameter Estimation} \label{Sec:estimation}

\subsubsection{Estimating the Source Density}

The energy spectra for five of the six sources (both with and without shielding) was provided in the form of an intensity histogram (count rate per energy level) with 2 keV bin widths and the first bin starting at 11 keV. We converted the intensity histograms into a probability mass function (pmf) at energy levels between 11 and 4001 by increments of 0.5 keV using frequency histograms (i.e., linear interpolation) and rescaling so it sums to one. 

We were not provided the energy spectra for source 6, but told it was "a combination of 99mTc and HEU" \cite{topcoder}. We assumed it was a 50-50 mix and used
\eq{
\hat{f}_6(x) = 0.5 \hat{f}_1(x) + 0.5 \hat{f}_5(x)
\label{eq:f6}
}
where $\hat{f}_k(x)$ is the estimated density (pmf) from source $k$. 
The estimated pmfs for all six densities are given in Fig.~\ref{fig:pmf}.

\subsubsection{Estimating the Null Density} 
The null density is estimated from the training runs with no source. An initial exploratory analysis suggested that the null density appeared stationary across each run with no discernible differences between runs. Thus, we estimated the null density from taking all energy measurements in a random 100 runs and applying kernel density estimation using a Gaussian kernel with bandwidth of 1 keV. The density estimate was then converted to a pmf by rescaling the density estimates at a range of energy levels between 11 and 4001 (keV) by increments of 0.5 keV to match the values from the six sources.

\subsubsection{Estimating $R_{ik}$}

\begin{figure}[t]
    \centering
    \includegraphics[width=.48\textwidth]{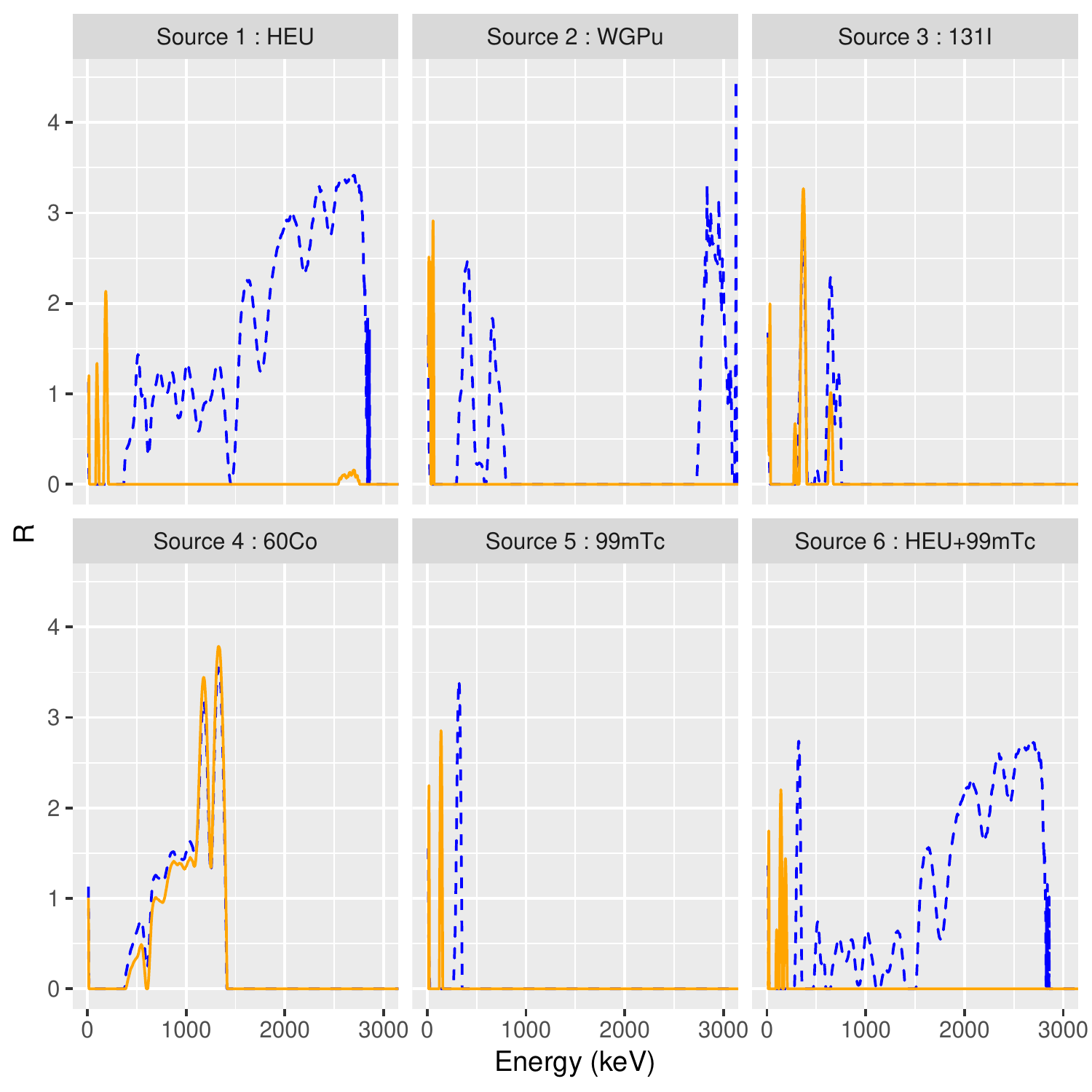}
    \caption{
    Estimated log density ratios, $R_k(x)$. The solid orange line is for unshielded and the dotted blue line is for shielded sources.
    }
    \label{fig:R}
\end{figure}

The value $R_{ik} = \max\{0, r_{ik}\}$ can be calculated in advance to enable a quick look up at run time. Because the null and source densities were estimated at the same set of discrete values, the log density ratios, $R_k(x) = \max(0, \log (f_k(x)/f_0(x)))$ can be pre-calculated for every possible energy level. See Figure~\ref{fig:R} for the estimated values of $R_k(x)$. 

\subsubsection{Estimating $S(k,\tau, h)$}

The value $S(k, \tau, h)$ is the smoothed value of $R$ over time. Because the source strength and run configurations (e.g., sensor speed, building interference, etc.) can affect the distance that particles are received by the sensor, we estimate it using a range of bandwidths $h \in H = \{0.5, 0.75, 1, 1.25, 1.5\}$.

\subsubsection{Estimating $Z(k, h)$}

The value $Z(\tau, h)$ is obtained by standardizing $\tilde{S}(\tau, h)$ by subtracting the mean and dividing by the standard deviation estimated from null (no source) runs. 
We used a sample of 900 null runs to estimate the mean, $\mu_0(k,h)$, and standard deviation, $\sigma_0(k,h)$ for each source $k \in K$ and bandwidth $h \in H$.

\section{Results}

Our method was applied to 8800 labeled runs. These runs comprised 4000 runs without an extraneous source (null runs) and 800 runs from each of the six sources. It is not known if the sources are shielded, however the time when the sensor is closest to the source is available. 
We evaluate our method on its ability to \emph{detect}, \emph{identify}, and \emph{localize}.

\subsection{Detection}

A source is declared to be present whenever the scan statistic, given in \eqref{eq:scanstat}, exceeds a threshold, i.e., $T \geq \phi$. The optimal threshold $\phi$ depends on the costs for false alarms and missed detections. Because this information varies across applications, we present results across a variety of thresholds. 

\begin{figure}
    \centering
    \includegraphics[width=.35\textwidth]{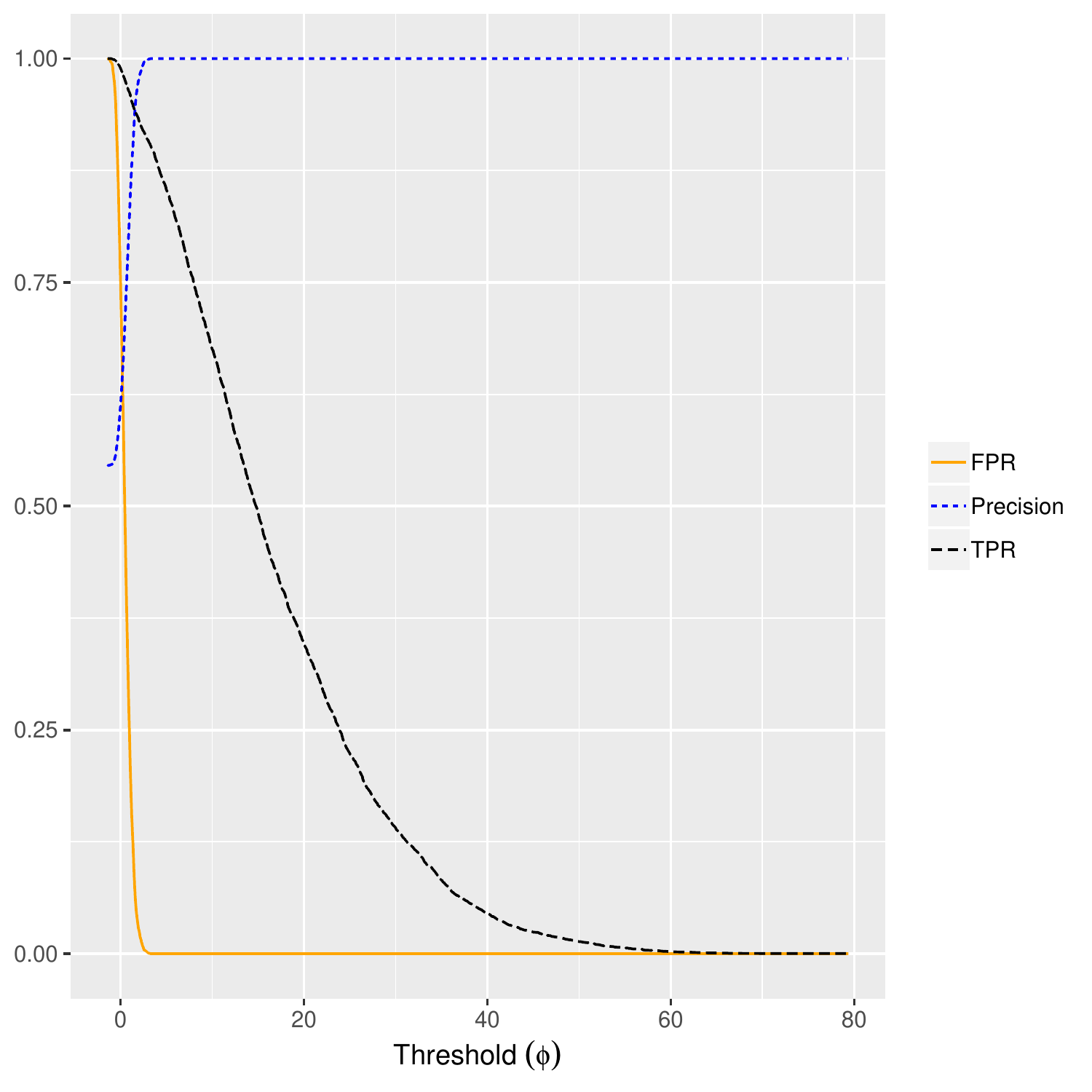}
    \caption{Performance of the method as a function of the decision threshold $\phi$. The solid orange line is the False Positive Rate (FPR), the dotted blue line is the Precision, and the dashed black line is the True Positive Rate (TPR). }
    \label{fig:metrics}
\end{figure}

Figure~\ref{fig:eval_Zdens} shows the distribution of $T$ under different sources (including the null). As expected, the test statistic is small for the null runs and usually large for the runs with an radiological source. Some common binary classification metrics are given in Figure~\ref{fig:metrics}; these metrics are used to construct ROC and Precision-Recall curves.  Because there are no null runs with $T \geq 3.26$,  the false positive rate (FPR) goes to zero after this value. The true positive rate decreases as the threshold is increased. The precision starts at $0.545$ ($4800/8800$),  since 54.5\% of the runs contain a source, and then quickly rises as the percentage of runs declared to have a source that actually have a source increases.   

\begin{figure}
    \centering
    \includegraphics[width=.48\textwidth]{fig_eval_Zdens}
    \caption{Histograms of the test statistic, $T$, under different sources.  }
    \label{fig:eval_Zdens}
\end{figure}

\subsection{Identification}

\begin{figure}
    \centering
    \includegraphics[width=.35\textwidth]{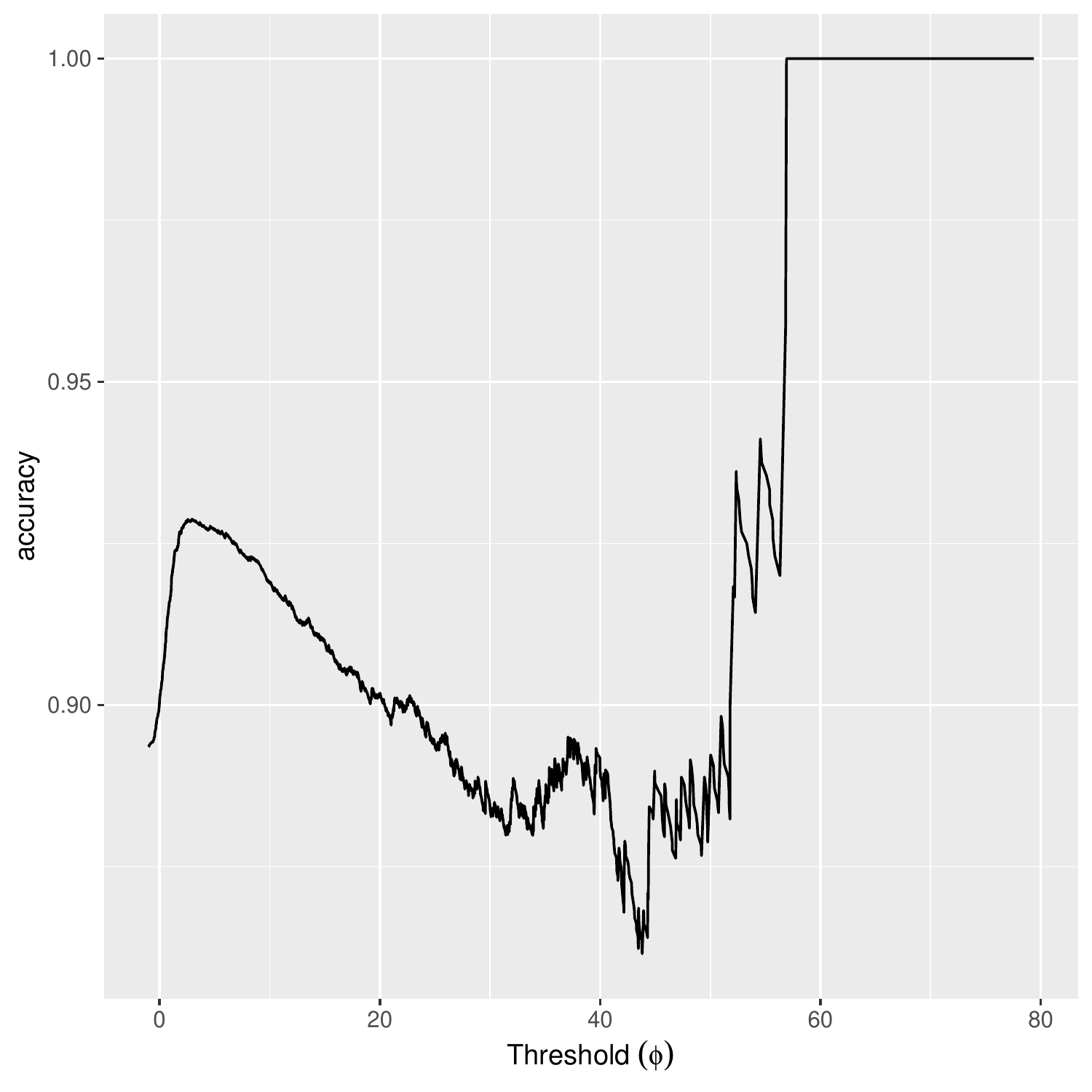}
    \caption{Identification accuracy as a function of the decision threshold $\phi$. This shows the proportion of sources correctly identified (i.e., $\hat{k} = k$) over all runs with an extraneous source. }
    \label{fig:accuracy}
\end{figure}

The ability of our method to correctly identify the source is also analyzed. Figure~\ref{fig:accuracy} gives the source identification accuracy as a function of the decision threshold. Notice that the performance is perfect when the evidence is strongest (i.e., $\phi> 50$) and never dips below 86.2\%. 

Table~\ref{tab:confusion} gives the confusion matrix obtained when $\phi = 2.5$ (selected arbitrarily). At this threshold, 3977 null runs are correctly assigned to have no source (true negatives) but 388 runs with a source are incorrectly classified as coming from the background (false negatives) and 23 of the null runs are incorrectly classified as coming from an extraneous source (false positives). 
The confusion matrix also reveals the difficulty of correctly discriminating source $k=6$; 60 runs get assigned incorrectly to source $k=1$ while 249 are incorrectly assigned to source $k=5$. This indicates potential improvements in the source identification with more knowledge of the composition of source 6 (see \eqref{eq:f6} for the current method of estimating the density of this source).  

\begin{table}
    \centering
        \caption{Confusion Matrix for $\phi = 2.5$.}
    \label{tab:confusion}
    
\begin{tabular}{ cc ccccccc|c}
\multicolumn{2}{c}{} & \multicolumn{7}{c}{Estimated Source}  \\
\multicolumn{2}{c}{} & 0 & 1 & 2 & 3 & 4 & 5 & 6  & Total      \\
    \cmidrule(lr){3-9}
\multirow{7}{*}{\STAB{\rotatebox[origin=c]{90}{True Source}}}
& 0 & 3977 & 3 & 7 & 1 & 3 & 4 & 5 & 4000\\
& 1 & 78 & 721 & 0 & 1 & 0 & 0 & 0 & 800\\
& 2 & 91 & 1 & 704 & 1 & 1 & 0 & 2 & 800\\
& 3 & 82 & 0 & 0 & 718 & 0 & 0 & 0 & 800\\
& 4 & 55 & 0 & 0 & 0 & 745 & 0 & 0 & 800\\
& 5 & 65 & 0 & 0 & 0 & 0 & 734 & 1 & 800\\
& 6 & 17 & 60 & 0 & 0 & 0 & 249 & 474 & 800 \\
    \cmidrule(lr){3-9}
& Total & 4365 & 785 & 711 & 721 & 749 & 987 & 482 & 8800\\    
\end{tabular}
\end{table}

\subsection{Localization}

\begin{figure}
    \centering
    \includegraphics[width=.35\textwidth]{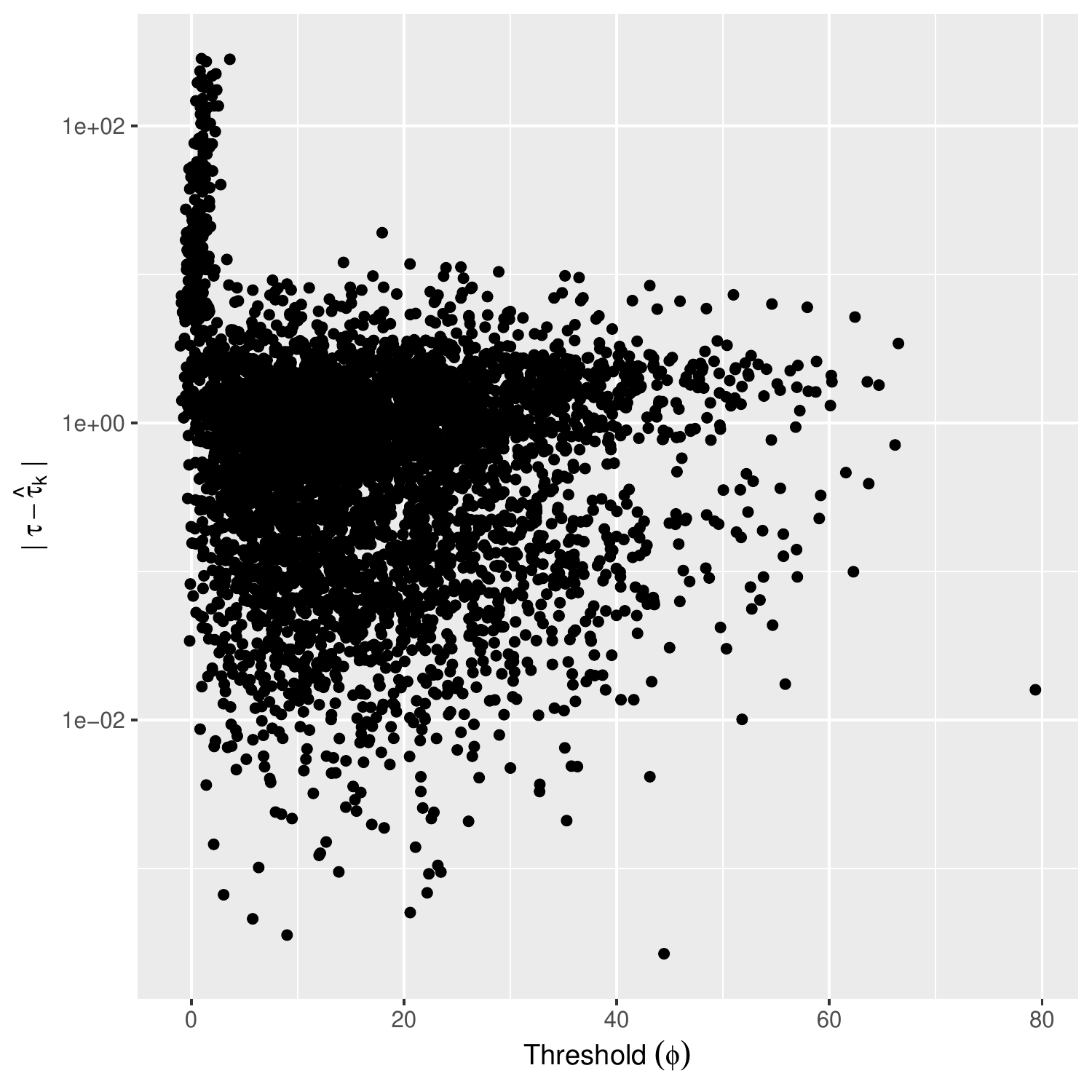}
    \caption{Localization performance as a function of the decision threshold $\phi$. This shows the \emph{distance},  $| \tau - \hat{\tau}_{\hat{k}} |$, between the true and estimated time when the sensor is closest to the source.}
    \label{fig:localization}
\end{figure}

Our method is also able to accurately estimate the time when the sensor is close to the source. Figure~\ref{fig:localization} shows the distance (in seconds) from estimated source location $\hat{\tau}_{\hat{k}}$ to the true location $\tau$ for all the runs with a source present. Using again a threshold of $\phi = 2.5$, we find that the median distance is 0.71 seconds, the mean distance is 1.13 seconds, and 95\% of distances are less than 3.05 seconds. 
Ideally, we would examine the spatial distance (rather then temporal) between the actual source and our estimate. Unfortunately, this information was not provided; the only related information provided is that the sensor speed can vary between 1 and 13.4 meters per second. Thus we can only generally say that our method can usually localize the source to within 40 meters.

\section{Related Work}

There have a been a few reported methods for radiation detection with mobile sensors in the literature. 
To locate the change in position of a radiological source that moves throughout a room indoors, \cite{Howse_leastsquares} 
developed a nonlinear state estimation algorithm that uses a constrained, feasible path, sequential quadratic program to solve a recursive least squares optimization problem. 
Several methods to detect a source are described in \cite{Chandy_modelsand} as well as an analysis of the degradation in detection ability due to mobile (and other time restricted) sensing platforms and sensitivity to the decision thresholds. 
A comparison of both frequentist and Bayesian approaches to detecting multiple sources is given in \cite{DBLP:conf/fusion/MorelandeRG07}. The authors find that a properly constructed Bayesian model performs best at identifying two and three sources. 
A comparison of quick deterministic methods with the flexibility of probabilistic methods is provided in \cite{DBLP:conf/fusion/LiuBC}. The authors find that a careful combination of the two approaches can provide a noticeable improvement in performance while significantly reducing the necessary computation. 
The Poisson-Clutter Split (PCS) algorithm combines a Poisson distribution model (for background plus source) with a Generalized Likelihood Ratio Test (GLRT) to detect a specific source \cite{shokhirev2012enhanced}. 

When multiple sources are possible, Gaussian mixture models can be used to detect and localize \cite{morelande2009radiation}.
A distributed sensor network is developed in \cite{liu2010analysis} to detection, identify, and locate a radiological source in large areas. 
 A network of small inexpensive mobile sensors was used to detect and identify sources in \cite{sullivan2016radioactive}. 
The mean-shift clustering algorithm was used in \cite{yang2010mobile} to detect the presence of multiple sources in large urban areas with a collection of many inexpensive mobile (e.g., 1500 cab mounted) sensors.

\section{Conclusions}

We have presented a method, based on scan statistics, to detect, identify, and localize illicit radiological material using mobile sensors in an urban environment. Our method handles varying levels of background radiation that changes according to the (unknown) environment. We can accurately determine if a source is present along a run as well as identify which of six possible sources generated the radiation. Our method can also localize the source, when detected, to within a few seconds. We have presented our results across a range of decision thresholds allowing stakeholders to evaluate the performance at different false alarm rates. 

Due to the simplicity of our approach, our models can be trained in a few minutes with very little training data and holds the potential to accurately detect, identify, and locate illicit sources in real-time.

\bibliographystyle{IEEEtran}
\bibliography{refs}

\end{document}